\documentclass[runningheads]{cl2emult}

\usepackage{makeidx}  
\usepackage{graphicx} 
\usepackage{subeqnar} 
\usepackage{multicol} 
\usepackage{cropmark} 
\usepackage{eso}      
\makeindex            

\begin{document}
\title*{Parameterisation of Galaxy Spectra in the 
\protect\newline2dF Galaxy Redshift Survey}
\toctitle{Parameterisation of Galaxy Spectra in the
\protect\newline 2dF Galaxy Redshift Survey}
%
%
\titlerunning{Parameterisation of Galaxy Spectra in the 2dFGRS}
%
\author{Darren Madgwick\inst{1}
\and Ofer Lahav\inst{1}
\and Keith Taylor\inst{2,3}
\and the 2dFGRS Team}
\authorrunning{Darren Madgwick et al.}
%
%
\institute{Institute of Astronomy, University of Cambridge, 
Madingley Road, Cambridge CB3 0HA, UK
\and Anglo-Australian Observatory, P.O Box 296, Epping, NSW 2121,
     Australia
\and Department of Astronomy, California Institute of Technology,
     Pasadena, CA 91125, USA}

\maketitle              

\begin{abstract}
The 2dF Galaxy Redshift Survey has already yielded
over 135,000 galaxy spectra (out of the planned 250,000). We present a
method
for the parameterisation of galaxy spectra in the 2dF and other large
redshift surveys using a new continuous parameter
$\eta=0.44pc_1-pc_2$, based on a 
Principal Components Analysis (where $pc_1$ and $pc_2$ are the
projections onto the first
and second principal components respectively).  This parameter is
designed to be 
robust to instrumental uncertainties and it represents the
absorption/emission strength of a galaxy.  
\end{abstract}

\section{The 2dF Galaxy Redshift Survey}

The 2dF Galaxy Redshift Survey\cite{ref1.1} is a combined
UK-Australian effort to 
map the distribution of galaxies brighter than $b_J=19.45$ (with
median redshift $z\simeq0.1$).
In so doing it will obtain a total of 250,000 galaxy spectra from
which redshifts will be determined, a factor of 10 more than any
previous survey.

Crucial to extracting the full scientific content of the survey is
an accurate characterisation of the galaxy population in terms of
spectral features. To
this end we address the problem of obtaining an accurate
and meaningful spectral classification that can be used in future
analyses e.g. of luminosity functions and correlation functions.

\section{Principal Components Analysis}

Principal Components Analysis (PCA) allows us to easily visualise
overall trends (if any)  within 
a multi-dimensional population~\cite{murt}~\cite{conoly}~\cite{folkes}.  It does this by
reducing the 
dimensionality of the parameter space (here spectral bins) down to
just those components which are the most informative 
(in the sense of maximum variance).
Any clustering in the space defined by the PCA is indicative of distinct
subpopulations within the sample.

In terms of reduced dimensionality we find that when we apply PCA to
the first 100,000 galaxy spectra obtained in the 2dFGRS that rather
than using the original 738 spectral 
channels to describe each spectrum, we can use only three projections
and still retain two thirds of the total variance within
the population~\cite{folkes}. The significance of each successive principal
component drops off very sharply so these first few components are by
far the most important.

\subsection{PCA on Volume-Limited Samples}

Due to the large number of galaxy spectra acquired so far in the
2dFGRS it 
has now become possible to consider volume-limited subsets of the
galaxies 
for the PCA.  This approach is more physically motivated
compared to considering the entire flux-limited data set in which the
spectral population can 
be biased by relationships between magnitude and spectral type.
  
\begin{figure}
\centering
\includegraphics[width=.6\textwidth]{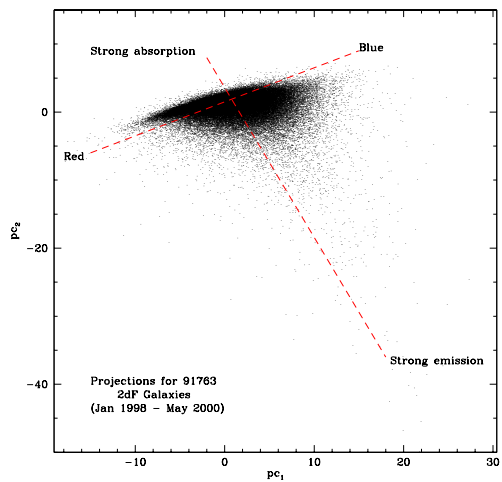}
\caption[]{Projections in the $pc_1$,$pc_2$ space for the galaxies
observed to date in the 2dFGRS.  The trends indicated have been
derived from the eigenspectra by identifying linear combinations that
either maximise or minimise the influence of the absorption/emission
features.}
\label{eps1}
\end{figure}

We find that using an absolute $B_j$ magnitude cut of -17.0 (corresponding
to a redshift $z\sim0.06$) gives a representative subset of the
local population.  We therefore perform the PCA on this subset of the
population to determine the principal components which are then used to
determine projections for the remaining galaxies (Fig.~\ref{eps1}).

\subsection{Identifying Trends}

\begin{figure}
\centering
\includegraphics[width=.6\textwidth]{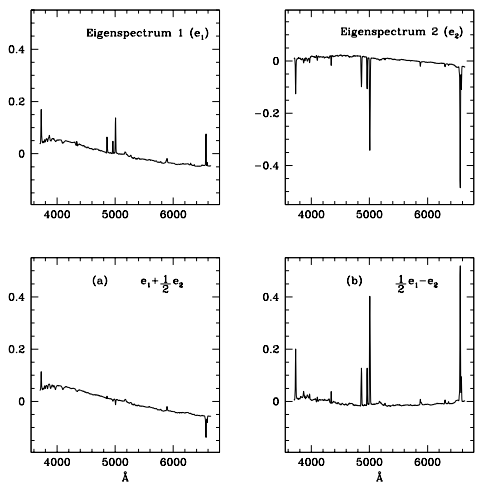}
\caption[]{The first two eigenspectra are shown in the top two panels
and the linear combinations which either minimise (a) or maximise (b)
the effect of emission/absorption features are shown in the bottom two.}
\label{eps2}
\end{figure}

 We can identify general trends in the distribution of projections
shown in Fig.~\ref{eps1} by 
considering the physical significance of the first two eigenspectra
(Fig.~\ref{eps2}).
It can be clearly
seen that whilst the first eigenspectrum contains both a continuum
component and a measure of the average emission/absorption strength, the
second contains only the latter (to a first approximation).  Therefore
by taking certain linear combinations of the two we can either
maximise or minimise the contribution from the line features in the
spectrum. By performing such an analysis we can conclude that
the distribution of Fig.~\ref{eps1} is indicative of a colour
sequence from red (bottom left) to blue (top right) and an orthogonal
sequence of absorption (top left) to emission (bottom right).

\section{Instrumental Uncertainties}

Unfortunately whilst carrying out observations using the 2dF
instrument allows us to collect an unprecedented number of spectra
simultaneously and hence determine many redshifts in a small amount of
time, we must make a
compromise with our spectral quality.  This is deemed to be a
necessary sacrifice in view of carrying out a project as
ambitious as the 2dFGRS, however if one wishes to characterise the
resultant population of galaxies using their spectral properties care
must be taken in order to ensure that the classification scheme is
robust to instrumental uncertainties.  

The 2dF instrument makes use of
up to 400 fibre optic cables with a diameter corresponding to
(depending on plate position) $\sim$2.0'' on the sky~\cite{taylor}.
The quality of the spectra can 
be compromised in many ways, the most important of which are:

\begin{itemize}
\item Astrometric errors in determining the actual position of the
galaxy on the sky can result in the fibre being placed in a position
that isn't optimal for light collection.  In general the position of a
galaxy centre is known to within $\sim0.4$''.  In addition to this
there is a random positioning error of the order of $\sim0.1$'' RMS.

\item The observed colour (continuum slope) of the galaxy depends on
the positioning of the fibre aperture.  This is due to small errors in
the 2dF corrector which result in 
relative spatial displacements between different components of a
galaxies spectrum and can be as
large as $>1$'' (at radius of $30$') between the incident blue and red
components of the observed spectrum.  

\item The apparent size of a galaxy on the sky can be larger than
the fibre aperture (which corresponds to $\sim2.5h^{-1}$ kpc at
the surveys median redshift of $z=0.1$ in an Einstein - de Sitter
Universe), the positioning  
of this aperture can therefore result in an unrepresentative spectrum
due to colour gradients within the galaxy.
\end{itemize}

Whilst the effect of these errors at first seems large one
must bear in mind that the average seeing at the 2dF is of the order of
$\sim1.5$'', enough to significantly smooth out most of the errors.  The
distortions mentioned in the second point 
result in unstable continuum measurements for the observed galaxies.

\section{Spectral Classification}

We are presented with several options in attempting to derive a
classification for the the observed 2dF galaxies based upon their
spectra.  However there are several issues that one must first consider;

\begin{itemize}
\item The distribution of galaxies in the ($pc_1$,$pc_2$) plane is
smooth, representing a continuous sequence from absorption to emission
and from red to blue.  There is no evidence to suggest any division into
distinct classes of galaxy spectra.

\item The instrumental errors result in unstable
continuum measurements. On the other hand small scale features such as
line-strengths are relatively unaffected.

\item The relationship between morphology and spectral properties is
not very well understood and hence attempting to anchor the
classification using a 
morphological training set would be premature.  A
more robust and quantifiable measure of spectral properties is
required.
\end{itemize}

\subsection{$\eta$ Parameterisation}

By projecting the $pc_1$ and $pc_2$ components of each galaxy onto the
linear combination which maximises the effect of emission/absorption
features we are in effect high-pass filtering the spectra.  Thus we
would expect (and indeed we find) this projection to be relatively stable
to uncertainties in continuum measurements.

By using this projection we are determining a measure of the average
emission/absorption strength of a galaxy which is easily quantifiable
and robust. In addition this projection is also representative of the
spectral sequence of the galaxy population since it is composed of
the most significant principal components. 

We therefore choose to adopt this projection, which we shall denote
$\eta$, as our continuous
measure of spectral type:

\begin{equation}
\eta = a.pc_1 - pc_2
\end{equation}

The value of $a$ which maximises the emission/absorption features is
found to be $a = 0.44 \pm 0.06$.

\begin{figure}
\centering
\includegraphics[width=.6\textwidth]{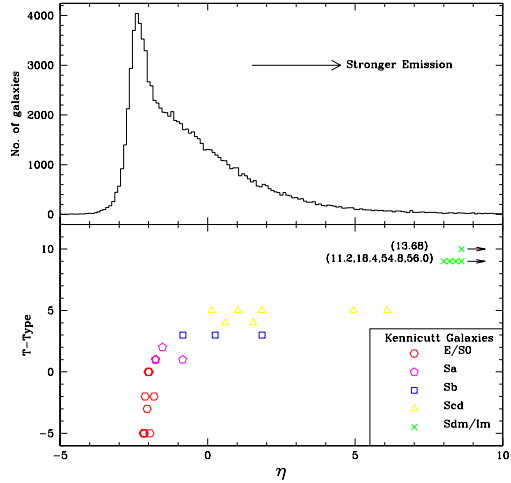}
\caption[]{Histogram (top) of the $\eta=0.44pc_1-pc_2$ projections for
the galaxies 
observed to date in the 2dFGRS.  Also shown (below) is the
$\eta$-morphology relation for a sample of galaxies from the  Kennicutt
Atlas~\cite{keni}.}
\label{eps3}
\end{figure}

In Fig.~\ref{eps3} the distribution of the $\eta$ projections are
shown for the galaxies observed to date in the 2dFGRS. Also shown in
the same figure is
the $\eta$-morphology relation for a sample of galaxies from the
Kennicutt Atlas~\cite{keni}.  Comparing the two data sets shows that
there is a correspondence between the sequence of $\eta$ and that of
morphology. Note that this correspondence can only be treated as an
approximation since our sample of Kennicutt galaxies is not complete
and only represents a few high Signal-to-Noise spectra with well determined
morphologies.

\section{Summary}

We have attempted to derive a meaningful and robust parameterisation 
of the observed spectral distribution in order to characterise the
galaxy population of the 2dFGRS.  
In so doing we have defined a new variable $\eta$ which represents the
most significant component of the galaxy sequence that is
unaffected by instrumental uncertainties, but still retains physical
information. 

The parameter $\eta$ is easily interpreted as an averaged
absorption/emission 
strength and can be shown to have significant correlations with more
conventional classifiers such as morphology and the H$\alpha$
equivalent width. 

There are certain limitations to our analysis, the most notable of
which is that we have only derived a classification in the broadest
sense. We have not attempted to identify the distinct sub-populations
that are known to exist buried within the galaxy sequence, however
there is no reason that such an analysis cannot be performed on the
data in the future.  Other significant limitations to our analysis 
are the assumption that each spectrum can be reconstructed using a
linear combination of only a few of the derived eigenspectra
and the restriction to variance as a measure of information (both of
which are implicitly assumed in the PCA).  These assumptions 
greatly simplify the analysis but are known to not be true
in general.  It is in fact possible to extend the analysis without the
use of these assumptions to perform an Independent Component Analysis.
Another example of a non-linear method that has recently been
investigated is that 
of an information-based clustering algorithm~\cite{slonim}.
Preliminary results from such analyses look promising.

\clearpage
\addcontentsline{toc}{section}{Index}
\flushbottom
\printindex

\end{document}